\documentclass{snmp2003}

\begin{document}

\FirstPageHeading{YourLastName}

\ShortArticleName{On Symmetries in (Anti)Causal (Non)Abelian Quantum Theories} 

\ArticleName{On Symmetries in (Anti)Causal (Non)Abelian\\
Quantum Theories}

\Author{Frieder KLEEFELD~$^\dag$}
\AuthorNameForHeading{F. Kleefeld}
\AuthorNameForContents{KLEEFELD F.}
\ArticleNameForContents{On Symmetries in (Anti)Causal (Non)Abelian Quantum Theories}

\Address{$^\dag$~CFIF, Instituto Superior T\'{e}cnico, Av. Rovisco Pais, P-1049-001 LISBOA, Portugal}
\EmailD{kleefeld@cfif.ist.utl.pt}


\Abstract{(Anti)causal boundary conditions being imposed on the (seemingly) Hermitian Quantum Theory (HQT) as described in standard textbooks lead to an (Anti)Causal Quantum Theory ((A)CQT) with an indefinite metric (see e.g.\ \cite{Kleefeld:2002au,Kleefeld:2002gw,Kleefeld:2001xd,Kleefeld:1998yj,Kleefeld:1998dg,Kleefeld:thesis1999}). Therefore, an (anti)causal neutral scalar field is not Hermitian, as one (anti)causal neutral scalar field consists of a non-Hermitian linear combination of two (Hermitian) acausal fields. Fundamental symmetries in (A)CQT are addressed. The quantum theoretical (transition) probability and antiparticle concepts are revised. Imaginary parts of cross sections and refraction indices are related.}

\section{Introduction to the idea of (anti)causal fields}
Textbooks (e.g.\ \cite{Kleefeld:Bjorken:dk}) on Quantum Field Theory (QFT) declare the neutral scalar Klein-Gordon (KG) field to be a {\em Hermitian} ({\em shadow} \cite{Kleefeld:Nakanishi:wx,Kleefeld:Stapp:1973aa}) {\em field} $\varphi(x)=\varphi^+(x)$ with real mass $m=m^\ast$ and Lagrangean ${\cal L}^{\,0}_{\,\varphi} (x) = \frac{1}{2}\left( (\partial\varphi (x) )^2  - \, m^2 \, \varphi (x)^2 \right)$. Note that $\varphi(x)$ is representing {\em one} (real) field-theoretical degree of freedom! Its equation equation of motion is \mbox{$(\partial^2 + m^2)\, \varphi (x) = 0$} yielding --- strictly speaking --- a principal value propagator \cite{Kleefeld:Stapp:1973aa} $\mbox{P} \,\frac{i}{p^2-m^2}$. Causality is typically enforced {\em afterwards} by the use of {\em causal} or {\em anticausal} Feynman propagators ($i/(p^2-m^2+i\,\varepsilon)$ (causal), $i/(p^2-m^2-i\,\varepsilon)$ (anticausal)) corresponding --- strictly speaking --- to the {\em causal KG equation} $(\partial^2 + m^2 - i\,\varepsilon )\, \phi (x)\, = 0$ or {\em anticausal KG equation} $(\partial^2 + m^2 + i\,\varepsilon )\, \phi^+(x) = 0$, respectively \cite{Kleefeld:Nakanishi:wx,Kleefeld:Nakanishi:1972pt,Kleefeld:2002au,Kleefeld:2002gw,Kleefeld:2001xd}.
The {\em causal KG field} $\phi(x)=(\varphi_1(x)+i\,\varphi_2(x))/\sqrt{2}$ and the {\em anticausal KG field} $\phi^+(x)=(\varphi_1(x)-i\,\varphi_2(x))/\sqrt{2}$ are {\em non-Hermitian} and represented by {\em two Hermitian shadow fields}  $\varphi_j(x)=\varphi_j^+(x)$ ($j=1,2$) yielding {\em two} (real) field-theoretical degrees of freedom. I.e.\ imposing causal boundary conditions on Quantum Theory (QT) leads (already at zero temperature) to a {\em doubling} of degrees of freedom like in Thermal Field Theory \cite{Kleefeld:Weldon:1998yk} or Open Quantum Systems \cite{Kleefeld:Romano:2003ms}. 
The non-Hermitian nature of QT should not surprise\footnote{In-fields and out-fields fulfil same {\em causal} KG equations: $(\partial^2 + m^2 - i\,\varepsilon ) \, \phi_{in} (x) = 0$, $(\partial^2 + m^2 - i\,\varepsilon ) \,\phi_{out} (x) = 0$. Hence, the Hilbert space of out-states is not obtained from the Hilbert space of in-states by Hermitian conjugation. Therefore we claim that in Quantum Mechanics (QM) ``bra's'' (in our notation: $\left<\!\left<\ldots\right|\right.\!$) are not obtained by Hermitian conjugation from ``ket's'' (in our notation: $\left|\ldots\right> = \left<\ldots \right|^+$)! An unexpected result in (A)CQT is obtained when considering the causal/anticausal KG, Dirac and Schr\"odinger equations for a complex mass $M=m-\frac{i}{2}\Gamma\simeq -i\,\varepsilon$.
In deriving the standard continuity equations as described in text books, i.e.
\begin{eqnarray} \makebox[1.8cm][l]{KG:} & & \partial_\mu [ \, \phi^+ (x) \; \partial^\mu \phi (x)  -  \phi (x) \; \partial^\mu \phi^+ (x) \, ]  = 2\, i \, \varepsilon \,\phi^+ (x) \, \phi (x) \; , \nonumber \\[2mm]
\makebox[1.8cm][l]{Dirac:} & & \partial_\mu [ \, i\, \bar{\psi} (x) \; \gamma^\mu\;  \psi (x) \, ] = -\, 2\, i \, \varepsilon \,\bar{\psi} (x) \, \psi (x) \; , \nonumber \\ 
\makebox[1.8cm][l]{Schr\"odinger:} & & i \partial_t \, [ \, \psi^+ (x) \, \psi (x) \, ] + \frac{1}{2\, m} \!\stackrel{\rightarrow}{\nabla} \!\cdot \,[ \, \psi^+ (x) \stackrel{\rightarrow}{\nabla} \psi (x) - \psi (x) \stackrel{\rightarrow}{\nabla} \psi^+ (x) \, ] = \nonumber \\
\makebox[1.8cm][l]{} & & =  \psi^+ (x) \, [ \, V(x) - V^+(x) \,] \, \psi (x) - \frac{i\,\varepsilon}{2\, m} \,[ \, \psi^+ (x) \stackrel{\rightarrow}{\nabla}  {}^2 \, \psi (x) + \psi (x) \stackrel{\rightarrow}{\nabla}  {}^2 \, \psi^+ (x) \, ] \; ,\nonumber
\end{eqnarray}
we have to observe that all these currents are not conserved due to the finite imaginary part of the mass (here $-\varepsilon$) or non-Hermitian {\em causal potentials} $V(x)$ being Laplace-transforms of {\em causal} propagators! The non-conservation of the Schr\"odinger current indicates a breakdown of the traditional probability interpretation of Schr\"odinger theory by Max Born in (A)CQT. Note that the ``massless'' causal Dirac equation $(i\!\not\! \partial + i \, \varepsilon ) \psi (x) = 0$ is not chiral invariant!}, but be taken into account!

\section{Introduction to the ``Nakanishi model''}
In the year 1972 N.\ Nakanishi \cite{Kleefeld:Nakanishi:wx,Kleefeld:Nakanishi:1972pt} investigated for curiosity the so called ``Complex-Ghost Relativistic Field Theory'', i.e.\ a theory for a KG field $\phi (x)$ with complex mass $M := m - \frac{i}{2} \, \Gamma$ (and the Hermitian conjugate field $\phi^+(x)$)\footnote{The formalism was later (1999-2000) independently rederived by the author (see e.g.\ \cite{Kleefeld:2002au,Kleefeld:2002gw,Kleefeld:2001xd}). N.\ Nakanishi called the non-Hermitian fields $\phi(x)$ and $\phi^+(x)$ ``Complex Ghosts''.}. In the following we want to introduce immediately isospin and to consider for convenience a set of $N$ equal complex mass KG fields $\phi_r (x)$ ($r = 1,\ldots,N$) (i.e.\ a charged ``Nakanishi field'' with isospin $\frac{N-1}{2}$) described by the ``Nakanishi Larangean''
\begin{eqnarray} {\cal L}_{0} (x) & = & \sum\limits_{r} \; \left\{
\frac{1}{2} \left( (\partial \phi_r (x) )^2  - \, M^2 \, \phi_r (x)^2 \right) \; + 
\frac{1}{2} \left( (\partial \phi_r^+ (x) )^2  - \, M^{\ast \, 2} \, \phi_r^+ (x)^2 \right) \right\} \; . \nonumber 
\end{eqnarray}
The Lagrange equations of motion for the causal and anticausal ``Nakanishi field'' $\phi_r (x)$ and $\phi_r^+ (x)$, i.e.\ $(\,\partial^2 + M^2) \,\phi_r (x) \; = \; 0$ and $(\,\partial^2 + M^{\ast \, 2}) \,\phi_r^+ (x) \; = \; 0$,  are solved by a Laplace-transform.\footnote{The result decomposing in ``positive'' \& ``negative'' complex frequencies (e.g.\ $\phi_r (x) = \phi_r^{(+)}(x) + \phi_r^{(-)}(x)$) is:
\begin{eqnarray}
\phi_r (x) & = & \int \frac{d^4p}{(2\pi)^3} \; \mbox{``}\,\delta (p^2 - M^2)\,\mbox{''} \; e^{-\, i \,  p\cdot x} \; a (p,r) \; \, = \int 
\frac{d^3 p}{(2\pi )^3 \; 2\, \omega \, (\vec{p}\,)}
\Big[ \, 
 a \, (\vec{p} , r ) \, e^{\displaystyle - \, i p x}  +
 c^+ (\vec{p} , r ) \, e^{\displaystyle i p x}
\,\Big] \Big|_{p^0 = \omega(\vec{p}\,)} \; , \nonumber \\
 & & \nonumber \\
\phi_r^+ (x) & = & \int \frac{d^4p}{(2\pi)^3} \, \mbox{``}\delta (p^2 - M^{\ast\,2})\mbox{''} \,e^{+\, i \,  p^\ast\cdot x} a^+ (p,r) = \int
\frac{d^3 p}{(2\pi )^3 2\omega^\ast (\vec{p}\,)}
\Big[ 
  c (\vec{p} , r ) \,e^{\displaystyle - i p^\ast x} +
 a^+ (\vec{p} , r ) \,e^{\displaystyle i p^\ast x}
\Big] \Big|_{p^0 = \omega(\vec{p})}, \nonumber
\end{eqnarray}
where we defined $a(\vec{p}\,):=a(p)|_{p^{0} = \omega (\vec{p}\,)}$ and $c^+(\vec{p}\,):=a(-p)|_{p^{0} = \omega (\vec{p}\,)}$ on the basis of the complex ``frequency'' $\omega(\vec{p}):=\sqrt{\vec{p}^2+M^{\,2}}$ ($\omega(\vec{0}):=M$). The meaning of the symbolic delta-distribution ''$\delta (p^2 - M^2)$'' for complex arguments has been illuminated by N.\ Nakanishi \cite{Kleefeld:Nakanishi1958,Kleefeld:Nakanishi:1972pt}. Nowadays it may be embedded in the framework of (tempered) Ultradistributions \cite{Kleefeld:Bollini:1998en}.} The ``Nakanishi model'' is quantized by claiming {\em Canonical equal-real-time commutation relations} \footnote{The standard Canonical conjugate momenta to the (anti)causal fields $\phi_r (x)$ and $\phi_r^+ (x)$ are given by $\Pi_r (x) := \delta \,{\cal L}_{0} (x)/\delta (\partial_0 \,\phi_r (x)) \stackrel{!}{=} \partial_0 \,\phi_r (x)$ and $\Pi_r^+ (x) := \delta \,{\cal L}_{0} (x)/\delta (\partial_0 \,\phi_r^+ (x)) \stackrel{!}{=} \partial_0 \,\phi_r^+ (x)$.}.
The non-vanishing commutation relations in configuration space are ($r,s = 1,\ldots,N$):
\begin{eqnarray}
{ [ \, \phi_r (\vec{x},t) , \Pi_s (\vec{y},t) \, ] }
 \; = \; i\, \delta^{\, 3} (\vec{x} - \vec{y}\,) \; \delta_{rs} \; , & &
{ [ \, \phi^+_r (\vec{x},t) , \Pi^+_s (\vec{y},t) \, ] }
 \; = \; i\, \delta^{\, 3} (\vec{x} - \vec{y}\,) \; \delta_{rs} \; . \nonumber
\end{eqnarray}
The resulting non-vanishing momentum-space commutation relations are ($r,s = 1,\ldots,N$):
\begin{eqnarray}
{ [ \, a \, (\vec{p},r) \; , \; c^+ (\vec{p}^{\,\prime},s) \, ] }
 & = &
 (2\pi)^3 \, 2 \; \omega \,(\vec{p}\,)\;\, \delta^{\, 3} (\vec{p} - \vec{p}^{\,\prime}\,) \; \delta_{rs} \; ,\nonumber \\
{ [ \, c \, (\vec{p},r) \; , \; a^+ (\vec{p}^{\,\prime},s) \, ] }
 & = &
 (2\pi)^3 \, 2 \; \omega^\ast(\vec{p}\,)\; \delta^{\, 3} (\vec{p} - \vec{p}^{\,\prime}\,) \; \delta_{rs} \; . \nonumber 
\end{eqnarray}
The Hamilton operator is derived by a standard Legendre transform (see e.g.\ \cite{Kleefeld:2001xd,Kleefeld:2002gw}):\footnote{$c^+$ are {\em creation operators} of Bosonic {\em particles}, while $a^+$ are {\em creation operators} of Bosonic {\em holes}. $c$ are {\em annihilation operators} of Bosonic {\em holes}, while $a$ are {\em annihilation operators} of Bosonic {\em particles}. The antiparticle/antihole concept will be sketched in section \ref{Kleefeld:section11}. (Anti)particles propagate towards the future, (anti)holes towards the past.} 
\begin{eqnarray} H_{0} & = &
\sum\limits_{r} \;\int\! \!
\frac{d^3 p}{(2\pi )^3 \; 2\, \omega \, (\vec{p}\,)}
\;
\frac{1}{2}\; \omega \,(\vec{p}\,)\; \Big( 
c^+ (\vec{p},r) \; a (\vec{p},r) +
a\, (\vec{p},r) \; c^+ (\vec{p},r) \; \Big) + \; \nonumber \\
 & + & \sum\limits_{r} \,\;\int\! \!
\frac{d^3 p}{(2\pi )^3 \; 2\, \omega^\ast (\vec{p}\,)}
\;
\frac{1}{2}\; \omega^\ast (\vec{p}\,)\; \Big( 
a^+ (\vec{p},r) \; c\, (\vec{p},r) +
c \, (\vec{p},r) \; a^+ (\vec{p},r) \; \Big) \; . \label{Kleefeld:EquationLabel1}
\end{eqnarray}
The ``Nakanishi-KG propagator'' is obtained by {\em real-time ordering} of causal KG fields \cite{Kleefeld:Nakanishi:wx,Kleefeld:2001xd,Kleefeld:Bollini:1998hj}:\footnote{For intermediate states with complex mass these propagators lead to {\em Poincar\'{e} covariant} results. At each interaction vertex coupling to {\em intermediate} complex mass fields there holds {\em exact} 4-momentum conservation. Only if complex mass fields with finite $\Gamma$ appeared as asymptotic states, then Poincar\'{e} covariance would be violated!}
\begin{eqnarray} \Delta_N (x-y) \;\;  \delta_{rs} 
 & := & -\,i\, \left<\!\left<0\right|\right.T\,[\,\phi_r\,(x)\, \phi_s\, (y)\,]\,\left|0\right>  \quad\;\;\;\; \;\;\, \stackrel{!}{=} \;\; \int\!\frac{d^{\,4}p}{(2\,\pi)^4}\;\; \frac{e^{-\,i\,p (x-y)}}{p^2 - M^2} \;\; \delta_{rs} \; . \label{Kleefeld:EquationLabel14}
\end{eqnarray}
The anticausal ``Nakanishi-KG propagator'' is obtained by Hermitian conjugation or by a vacuum expectation value of an {\em anti-real-time ordered} product of two anticausal fields.

\section{(Anti)causal Quantum Mechanics}
The representation independent, time-dependent Schr\"odinger equation is $ i \, \partial_t \, \left|\psi(t)\right> \; = \; H \, \left|\psi(t)\right>$. Its adjoint is given by $-\,i\, \partial_t \left<\!\left<\psi(t) \right|\right. \; = \; \left<\!\left<\psi(t) \right|\right. H$. 
In 1-dim.\ QM we consider Hamilton operator of the (anti)causal Harmonic Oscillator \cite{Kleefeld:Nakanishi:1972pt,Kleefeld:2002gw,Kleefeld:2001xd,Kleefeld:1998yj,Kleefeld:1998dg,Kleefeld:thesis1999,Kleefeld:Kossakowski2002} (see equation~(\ref{Kleefeld:EquationLabel1})) $H \; = \; H_C + H_A \; = \; \frac{1}{2} \, \omega \, [c^+, a\, ]_\pm + \frac{1}{2} \, \omega^\ast \, [a^+, c\, ]_\pm \; = \;  \omega \, (c^+ a \pm \frac{1}{2}) + \omega^\ast \, (a^+ c \pm \frac{1}{2})$ ($\pm$ for Bosons/Fermions\footnote{The Fermionic oscillator we tend to denote by $H = \frac{1}{2} \, \omega \, \{\,d^+, b\, \} + \frac{1}{2} \, \omega^\ast \, \{\,b^+, d\, \}$ with $\{\,b, d^+ \} = 1$ etc.\,.}) with\footnote{All other possible commutators (Bosons) or anticommutators (Fermions) of $a$, $c$, $a^+$, and $c^+$ vanish!}
\begin{eqnarray} \left( \begin{array}{cc} {[c,c^+]_\mp} & {[c,a^+]_\mp} \\
{[a,c^+]_\mp} & {[a,a^+]_\mp} \end{array}\right) \quad = \quad \left( \begin{array}{cc} 0 & 1 \\
1 & 0 \end{array}\right) \quad = \quad \mbox{``indefinite metric''} \; , \nonumber 
\end{eqnarray}
yielding $[H_C,H_A]=0$ \footnote{A ``Hermitian'' Hamilton operator quantized with a non-trivial (e.g.\ indefinite) metric is called {\em pseudo-Hermitian}! Pseudo-Hermiticity (\& pseudo-unitarity) presently promoted \cite{Kleefeld:Znojil2003} by M.\ Znojil and A.~Mostafazadeh. Ideas go back to names like W.\ Pauli, P.A.M.\ Dirac, S.N.\ Gupta, K.\ Bleuler, and E.C.G.\ Sudarshan.}. The right/left eigensystem of the Hamilton operator is found by diagonalizing the stationary Schr\"odinger equation $(H-E) \left|\psi\right> = 0$ \& its adjoint $\left<\!\left<\,\psi\right|\right.\! (H-E) = 0$.
The resulting (normalized) right eigenstates $\left|n,m\right>$ and left eigenstates $\left<\!\left<\,n,m\right|\right.$ for the eigenvalues $E_{n,m} = \omega  \, (n \pm \frac{1}{2}) + \omega^\ast \, (m \pm \frac{1}{2})$ are given by $\left|n,m\right> \; := \; \frac{1}{\sqrt{n!\,m!}} \,\, (c^+)^n (a^+)^m \left|0\right>$ and $\left<\!\left<\,n,m\right|\right. \; := \; \frac{1}{\sqrt{m!\,n!}} \, \left<\!\left<\,0\right|\right. c^m \, a^n$ (Bosons: $n,m\in\{0,1,2,\ldots\}$; Fermions: $n,m\in\{0,1\}$).
The (bi)orthogonal eigenstates are complete: $\left<\!\left<\right.\right.\!\!n^\prime,m^\prime\left|n,m\right> = \delta_{n^\prime n} \, \delta_{m^\prime m}$,
$\sum_{n,m} \left|n,m\right>\left<\!\left<\,n,m\right|\right. = \mbox{\bf 1}$. In holomorphic representation (see e.g.\ \cite{Kleefeld:Chruscinski2002}) the time-dependent Schr\"odinger equation and its adjoint are: 
\begin{eqnarray} i\, \partial_t \, \left<\!\left<z,z^\ast\right| \psi(t)\right> & = & \int dz^\prime dz^{\prime\ast} \, \left<\!\left<\right.\right.\!\!z,z^\ast | H |z^\prime,z^{\prime\ast}\!\left.\right> \, \left<\!\left<\right.\right.\!\!z^\prime,z^{\prime\ast}\left| \psi(t)\right> \; , \nonumber \\
-\,i\, \partial_t \left<\!\left<\psi(t)\right| z,z^\ast\right> & = & \int dz^\prime dz^{\prime\ast} \,\left<\!\left<\right.\right.\!\!\psi(t)|z^\prime,z^{\prime\ast}\!\left.\right>\, \left<\!\left<\right.\right.\!\!z^\prime,z^{\prime\ast} | H\!\left|z,z^\ast\right> \; . \nonumber
\end{eqnarray}
\noindent The holomorphic representation of the Hamilton operator ($H(z,z^\ast)=H_C(z)+H_A(z^\ast)$) is:
\begin{eqnarray} \lefteqn{H(z^\prime,z^{\prime\ast};z,z^\ast) \quad = \quad \left<\!\left<\right.\right.\!\!z^\prime,z^{\prime\ast} | H\!\left|z,z^\ast\right>\quad = \quad H(z,z^\ast) \left<\!\left<\right.\right.\!\!z^\prime,z^{\prime\ast} \! \left|z,z^\ast\right> \quad =}\nonumber \\ 
 & = & \left( -\, \frac{1}{2\, M} \frac{d^2}{dz^2} + \frac{1}{2} \; M \, \omega^2 \, z^2  \; -\, \frac{1}{2\, M^\ast} \frac{d^2}{dz^{\ast\,2}} + \frac{1}{2} \; M^\ast \, \omega^{\ast \,2} \, z^{\ast\,2} \right) \left<\!\left<\right.\right.\!\!z^\prime,z^{\prime\ast} \!\left|z,z^\ast\right> \; . \nonumber 
\end{eqnarray}
The resulting stationary Schr\"odinger equations in holomorphic representation are given by
$H(z,z^\ast) \;\left<\!\left<\,z,z^\ast \right|\right.\!\left.\psi\right> = E \; \left<\!\left<\,z,z^\ast\right|\right.\!\left.\psi\right>$, $\left<\!\left<\,\psi\,\right|\right.\!\left.z,z^\ast\right> \; H(z,z^\ast) = E \; \left<\!\left<\,\psi\,\right|\right.\!\left.z,z^\ast\right>$.\footnote{The eigensolutions for the eigenvalues $E_{n,m}$ are given by ($(M \omega)^{-1/2}=$ complex oscillator length):
\begin{eqnarray} \left<\!\left< z,z^\ast\right|\right.\left.\! n,m\right> & = & \quad\;\; i^{\,n+m}\, \sqrt{\frac{|M\omega|}{2^{\,n+m} \, n!\,  m! \, \pi}} \, \exp\left(-\,\frac{1}{2}\, ( M \,\omega \, z^2 + M^\ast \,\omega^\ast \, z^{\ast\,2})\right) \; H_n(z\; \sqrt{M \,\omega} \, ) H_m(z^\ast \; \sqrt{M^\ast \,\omega^\ast} \, ) \; , \nonumber \\[3mm]
\left<\!\left< n,m\right|\right.\left.\! z,z^\ast \right> & = & (-i)^{\,m+n}\, \sqrt{\frac{|M\omega|}{2^{\,m+n} \, m!\,  n! \, \pi}} \, \exp\left(-\,\frac{1}{2}\, (M^\ast \,\omega^\ast \, z^{\ast\,2}+M \,\omega \, z^2)\right) \; H_m(z^\ast \; \sqrt{M^\ast \,\omega^\ast} \, )  H_n(z\; \sqrt{M \,\omega} \, )\, \; . \nonumber
\end{eqnarray}}
Note the non-trivial identities $\int dz \, dz^\ast \; \left|z,z^\ast\right>\left<\!\left<\,z,z^\ast\right|\right. = \mbox{\bf 1}$, $\left<\!\left<\,z,z^\ast\right.\right.\!\!\left|z^\prime,z^{\prime\,\ast}\right> = \delta(z-z^\prime) \; \delta(z^\ast-z^{\prime\,\ast})$. 
\section{Lorentz transformations/covariance of (anti)causal systems}
A Lorentz transformation $\Lambda^{\mu}_{\,\;\nu}$ for a given metric $g_{\mu\nu}$ is defined by $\Lambda^{\mu}_{\,\;\rho} \; g_{\mu\nu} \; \Lambda^{\nu}_{\,\;\sigma} = g_{\rho\sigma}$. Let $n^{\mu}$ be a timelike unit 4-vector $(n^2 = 1)$ and $\xi^{\mu}$ an \underline{arbitrary complex} 4-vector with $\xi^2 \not= 0$. We want to construct \cite{Kleefeld:2002au,Kleefeld:2002gw,Kleefeld:2001xd} a Lorentz transformation $\Lambda^{\mu}_{\,\;\nu}(\xi)$ relating the 4-vector $\xi^{\mu}$ with its ``restframe'', i.e. $\xi^{\, \mu} = \Lambda^{\,\mu}_{\,\;\,\nu} (\xi) \; n^{\,\nu} \sqrt{\xi^2}$ and $n_{\,\nu} \sqrt{\xi^2} = \xi_{\, \mu} \, \Lambda^{\,\mu}_{\,\;\,\,\nu} (\xi)$. After defining the {\em inversion matrix} $P^{\,\mu}_{\,\;\,\nu} := 2\, n^{\,\mu} \, n_{\,\nu} - \, g^{\,\mu}_{\,\;\,\nu}$
we find {\em two ``symmetric''} and {\em two ``asymmetric''} matrices $\Lambda^{\mu}_{\,\;\nu}(\xi)$ solving the defining equation $\Lambda^{\mu}_{\,\;\rho} \; g_{\mu\nu} \; \Lambda^{\nu}_{\,\;\sigma} = g_{\rho\sigma}$, i.e. ($\xi \cdot n := \xi^{\,\mu} \, n_\mu$):
\begin{eqnarray} \Lambda^{\,\mu}_{\,\;\,\nu} (\xi) & = &  \pm \, \left\{ g^{\,\mu}_{\,\;\,\nu} - \, \frac{\sqrt{\xi^2}}{\sqrt{\xi^2} \mp \xi \cdot n} \, \Big[ n^{\,\mu} \mp \frac{\xi^{\,\mu}}{\sqrt{\xi^2}} \Big] \Big[ n_{\,\nu} \mp \frac{\xi_{\,\nu}}{\sqrt{\xi^2}} \Big] \, 
\right\} \; , \nonumber \\
\Lambda^{\,\mu}_{\,\;\,\nu} (\xi) & = &  \pm \, \left\{ g^{\,\mu}_{\,\;\,\rho} - \, \frac{\sqrt{\xi^2}}{\sqrt{\xi^2} \mp \xi \cdot n} \, \Big[ n^{\,\mu} \mp \frac{\xi^{\,\mu}}{\sqrt{\xi^2}} \Big] \Big[ n_{\,\rho} \mp \frac{\xi_{\,\rho}}{\sqrt{\xi^2}} \Big] \, 
\right\} \,  P^{\,\rho}_{\,\;\,\nu} \; . \nonumber
\end{eqnarray}
For the real Lorentz group the 4 solutions are related to the well known ortho-chronous/non-ortho-chronous proper/improper Lorentz transformations! Convince yourself that for the metric $+,-,-,-$ one of the ``asymmetric'' Lorentz boosts is given by: 
\begin{eqnarray} \Lambda^{\,\mu}_{\,\;\,\nu} (\xi) = 
\left( \begin{array}{cc} \frac{\xi^0}{\sqrt{\xi^2}} & \frac{\vec{\xi}^{\,T}}{\sqrt{\xi^2}} \\
 & \\
\frac{\vec{\xi}}{\sqrt{\xi^2}} & 1_3 + \frac{\vec{\xi} \; \vec{\xi}^{\,T}}{\sqrt{\xi^2} \, (\sqrt{\xi^2} + \xi^0)}
\end{array} \right) & \Rightarrow & \Lambda^{\,\mu}_{\,\;\,\nu} (p)|_{p^0=\omega(\vec{p})} = 
\left( \begin{array}{cc} \frac{\omega (\vec{p}\,)}{M} & \frac{\vec{p}^{\,T}}{M} \\
 & \\
\frac{\vec{p}}{M} & 1_3 + \frac{\vec{p} \; \vec{p}^{\,T}}{M \, (M + \omega (\vec{p}\,))}
\end{array} \right) . \nonumber 
\end{eqnarray}
In the right expression we chose $\xi^\mu=p^\mu$ with $p^2 = M^2$ and $M = m - i\, \frac{\Gamma}{2}$.
Some properties of $\Lambda^{\,\mu}_{\,\;\,\nu} (p)$ have already discussed in Refs.\ \cite{Kleefeld:2002au,Kleefeld:2002gw,Kleefeld:2001xd}.\footnote{G.P.\ Pron'ko \cite{Kleefeld:Pron'ko:1996be} could of course argue that such a Lorentz boost between $\vec{p}$ and $\vec{p}^{\;\prime}$ {\em ``$\ldots$ understood literary leads to nonsense because the transformed space components of the momentum become complex. $\ldots$''} Certainly this argument is {\em only} true for complex mass fields with finite $\Gamma$ being treated as {\em asymptotic states}. Yet --- as argued in the context of equation (\ref{Kleefeld:EquationLabel14}) --- for complex mass fields in {\em intermediate states} Poincar\'{e} invariance is completely restored! The crucial difference between traditional HQT and (A)CQT is that in HQT fields are claimed to be representations of the covering group of the real Lorentz group $L^\uparrow_+$, while in (A)CQT even asymptotic (anti)causal states are representations of the covering group of the complex Lorentz group $L_+(C)$ (or, more generally, the covering group of the respective Poincar\'{e} group) \cite{Kleefeld:Greenberg:2003nv}.}
\section{The (anti)causal Dirac theory}
The causal Dirac equation and its relatives obtained by Hermitian conjugation and/or transposition are \cite{Kleefeld:2002au,Kleefeld:2002gw,Kleefeld:1998yj,Kleefeld:1998dg,Kleefeld:thesis1999} $( i \! \stackrel{\;\,\rightarrow}{\partial\!\!\!/} - \, M ) \; \psi_r (x) = 0$, $( i \! \stackrel{\;\,\rightarrow}{\partial\!\!\!/} - \, \bar{M} ) \; \psi_r^c (x) = 0$, $\overline{\psi_r^c} (x) \; ( - \, i \! \stackrel{\;\,\leftarrow}{\partial\!\!\!/} - \, M ) = 0$, and $\bar{\psi}_r (x) \; ( - \, i \! \stackrel{\;\,\leftarrow}{\partial\!\!\!/} - \, \bar{M} ) = 0$ ($M := m - \frac{i}{2} \, \Gamma$, $\bar{M}:= \gamma_0 \, M^+ \gamma_0$). Note that $\;r=1, \ldots, N$ is an isospin index and $\psi_r(x)$, $\bar{\psi}_r (x)$, $\psi_r^c (x)= C \, \gamma_0 \,\psi_r^\ast (x)$, $\overline{\psi_r^c} (x)= \psi_r^T(x) \, C$ are Grassmann fields.
The underlying Lagrange density is given by \cite{Kleefeld:2002au,Kleefeld:2002gw,Kleefeld:1998yj,Kleefeld:1998dg,Kleefeld:thesis1999} ($N=1$ yields neutrinos!)
\begin{eqnarray} {\cal L}^{\,0}_{\,\psi} (x) & = & \sum\limits_r \,\frac{1}{2} \left( \; \overline{\psi_r^c} (x) \, ( \frac{1}{2} \, i \! \stackrel{\;\,\leftrightarrow}{\partial\!\!\!/} \! -  M ) \, \psi_r (x) \; + \; \bar{\psi}_r (x) \, ( \frac{1}{2} \, i \! \stackrel{\;\,\leftrightarrow}{\partial\!\!\!/} \! -  \bar{M} ) \, \psi_r^c (x) \; \right) \; . \nonumber
\end{eqnarray}
4-spinors $u(p,s)\equiv v(-p,s)$ in complex 4-momentum space ($s=\pm \, \frac{1}{2}$) are introduced by the defining equation $( \, p\!\!\!/ -  \sqrt{p^2} \; ) \; u(p,s) \;  = \; 0\; \Leftrightarrow \;  \overline{u^c}(p,s) \, ( \, - \, p\!\!\!/ -  \sqrt{p^2} \; ) \; = \; 0$.
The spinors are normalized according to $\mbox{sgn[Re}(p^0) \,] \; \sum_s \, u(p,s) \; \overline{v^c}(p, s) =  \; p\!\!\!/ +  \sqrt{p^2}$ for $\mbox{Re}[p^0]\;\ne\;0$. Equations of motions are solved by a Laplace-transformation.\footnote{The result decomposing in ``positive'' \& ``negative'' complex frequencies (e.g.\ $\psi_r (x) = \psi_r^{(+)}(x) + \psi_r^{(-)}(x)$) is:
\begin{eqnarray} \psi_r (x) & = & \sum\limits_s \int \frac{d^3p}{(2\pi)^3 \; 2 \, \omega (\vec{p}\,)} \;\; [ \; \,e^{-\, i \,  p\cdot x} \;\;\,\, b_{\,r} (p, s) \, u(p,s) \; \, + \, e^{+ \, i \,  p\cdot x} \; b_{\,r} (- p, s) \, v(p,s)\;] \big|_{p^0 = \omega (\vec{p}\,)} \; , \nonumber \\
 \psi_r^c (x) & = & \sum\limits_s \int \frac{d^3p}{(2\pi)^3 \; 2 \, \omega^\ast (\vec{p}\,)} \, [ \; \,e^{+\, i \,  p^\ast\cdot x} \; b^+_{\,r} (p, s) \, u^c(p,s) + \, e^{- \, i \,  p^\ast\cdot x} \; b^+_{\,r} (- p, s) \, v^c(p,s)\;] \big|_{p^0 = \omega (\vec{p}\,)} \; , \nonumber \\
 \overline{\psi_r} (x) & = & \sum\limits_s \int \frac{d^3p}{(2\pi)^3 \; 2 \, \omega^\ast (\vec{p}\,)} \, [ \; \,e^{+ i \,  p^\ast \cdot x} \;\, b^+_{\,r} (p, s) \, \bar{u}(p,s) \; \, + \, e^{- \, i \,  p^\ast\cdot x} \; b^+_{\,r} (- p, s) \, \bar{v}(p,s)\;] \big|_{p^0 = \omega (\vec{p}\,)} \; , \nonumber \\
 \overline{\psi^c_r} (x) & = & \sum\limits_s \int \frac{d^3p}{(2\pi)^3 \; 2 \, \omega (\vec{p}\,)} \, \, \;[ \; \,e^{- i \,  p \cdot x} \;\;\,\,  b_{\,r} (p, s) \, \overline{u^c}(p,s)\, + \, e^{+ \, i \,  p \cdot x} \; b_{\,r} (- p, s) \, \overline{v^c}(p,s)\;] \big|_{p^0 = \omega (\vec{p}\,)}\; . \nonumber 
\end{eqnarray}
Note that $d_r^+(p,s):= b_r(-p,s) \;$, $\{b_r(\vec{p},s),d_{r^\prime}^+(\vec{p}^{\,\prime},s^{\,\prime})\}= (2\pi)^3 \,2\omega (\vec{p}\,)\,\delta^3(\vec{p}-\vec{p}^{\,\prime}) \, \delta_{s s^\prime}\delta_{r r^\prime}, \ldots$.}
Note that the spinors obey the analyticity property $u^c(p,s) \,  = \, u(-\,p^\ast ,s) \,  = \, v(p^\ast ,s)$. 
The (anti)causal Dirac equation is Lorentz covariant due to standard transformation properties of spinors and $\gamma$-matrices, i.e.\ $u(p) = S(\Lambda(p)) \; u(\sqrt{p^2} \; n)$ and $S^{-1} (\Lambda (p)) \, \gamma^{\,\mu} \, S (\Lambda (p)) \, = \, \Lambda^{\,\mu}_{\,\,\;\nu} (p) \; \gamma^{\,\nu}$.
The causal ``Nakanishi-Dirac propagator'' of a spin 1/2 Fermion is obtained by {\em standard Fermionic real-time ordering} of causal Dirac fields:
\[S_{N} (x-y)_{\alpha\beta} \;\;  \delta_{rs} \; := \; - i \, \left<\!\left<0\right|\right. T\,[\,(\psi_r (x))_\alpha \, (\,\overline{\psi^c_s} (y))_\beta\,] \left|0\right>  \; \stackrel{!}{=} \; \int\!\frac{d^{\,4}p}{(2\,\pi)^4}\;\; \frac{e^{-i\,p (x-y)}}{p^2 - M^2} \, (\not\!p + M)_{\alpha \beta}\; \delta_{rs} \; . \]
The anticausal ``Nakanishi-Dirac propagator'' is obtained by Hermitian conjugation or by a vacuum expectation value of a {\em anti-real-time ordered} product of two anticausal Dirac fields.
\section{(Anti)causal massive and ``massless'' vector fields}
With\footnote{The problem of constructing (anti)causal vector fields is twofold. First one has to be aware that even a ``massless'' (anti)causal field has to be treated as if it were massive due to the at least infinitesimal imaginary part of its complex mass. That even non-Abelian massive vector fields can be treated consistently within QFT without relying on a Higgs mechanism has been shown by Jun-Chen Su in a renormalizable and unitary formalism \cite{Kleefeld:Su:1998wy}. Secondly, one has to be able to construct polarization vectors based on a boost of complex mass fields.}
$\vec{e}^{\,\,(i)}\; (i=x,y,z)$ and $\vec{e}^{\,\,(i)} \cdot \vec{e}^{\,\,(j)} = \delta^{ij}$ we define the polarization vectors
$\varepsilon^{\,\mu \, (i)} (p) := \Lambda^\mu_{\;\nu} (p) \; \varepsilon^{\,\nu \, (i)} (\sqrt{p^2},\vec{0}) = \Lambda^\mu_{\;\nu} (p) \, (0,\vec{e}^{\,\,(i)})^{\,\nu}$. In the chosen unitary gauge they obey $p^\mu  \, \varepsilon_\mu^{\,(i)} (p) = 0$, $\varepsilon^{\,\mu \, (i)} (p) \, \varepsilon_\mu^{\, (j)} (p) = - \,\delta^{ij}$, and $\sum_i \varepsilon^{\,\mu \, (i)} (p) \, \varepsilon^{\,\nu \, (i)} (p) =  - g^{\,\mu\nu} + \frac{p^{\mu}\, p^{\nu}}{p^2}$.
Based on these polarization vectors the Bosonic field operators for (anti)causal vector fields (respecting (anti)causal commutation relations) are easily introduced according to ($r=1,\ldots,N=$ isospin index)
\begin{eqnarray} V_r^\mu (x) & = & \sum_{j} \int \frac{d^3p}{(2\pi)^3 \; 2\, \omega (\vec{p}\,)} \;\;\varepsilon^{\,\mu \, (j)} (p) \; \, [ \; e^{-\, i \,  p\cdot x} \; a_r (p,j) + e^{+ \, i \,  p\cdot x} \; a_r (- \, p,j)] \Big|_{p^{0} = \omega (\vec{p}\,)\,} \nonumber \\
 (V_r^\mu (x))^+ & = & \sum_{j}\int \frac{d^3p}{(2\pi)^3 \; 2\, \omega^\ast (\vec{p}\,)} \;\varepsilon^{\,\mu \, (j)} (p^\ast) \, [ \; e^{+\, i \,  p^\ast\cdot x} \; a_r^+ (p,j) + e^{- \, i \,  p^\ast \cdot x} \; a_r^+ (- \, p,j)] \Big|_{p^{0} = \omega (\vec{p}\,)} \; . \nonumber 
\end{eqnarray}
\section{Norm conservation in QM, Klein-Gordon- and Dirac-theory}
(Anti)causal KG/Dirac fields are decomposed into positive/negative complex frequency parts (e.g.\ $\phi (x) = \phi^{(+)}(x) + \phi^{(-)}(x)$ and $\psi (x) = \psi^{(+)}(x) + \psi^{(-)}(x)$).
Subtraction of equations of motion for $\phi^{(\pm)} (x)$, $\psi^{(\pm)}$, $\psi (x)$ and respective adjoints\footnote{KG: $(\partial^2 +M^2) \, \phi^{(\pm)} (x) = 0$, $\phi^{(\pm)} (x) \,(\,\stackrel{\leftarrow}{\partial}{}^2 +M^2 ) \; = 0$; Dirac: $(i\!\not\!\partial - M ) \, \psi^{(\pm)} (x) = 0$, $\overline{\psi^{(\pm)c}} (x) \, (-\,i\!\stackrel{\leftarrow}{\not\!\partial} - \,M ) = 0$; Schr\"odinger: $+ \, i \, \partial_t \, \psi (x) \; = (- \, \frac{1}{2\,M} \, \stackrel{\rightarrow}{\nabla} {}^2 + V(x) ) \, \psi (x)$, $- \, i \, \tilde{\psi} (x) \stackrel{\leftarrow}{\partial}_t \, = \, \tilde{\psi} (x) \, (- \, \frac{1}{2\, M} \, \stackrel{\leftarrow}{\nabla} {}^2 + V(x) )$.} yields the following continuity equations:\\ 
\noindent KG: $\partial_\mu [ \, \phi^{(\mp)} (x) \; \partial^\mu \phi^{(\pm)} (x)  - (\partial^\mu \phi^{(\mp)} (x))\,\phi^{(\pm)} (x) \, ]  = 0$, Dirac: $\partial_\mu [ \, i\; \overline{\psi^{(\mp)c}} (x) \; \gamma^\mu\;  \psi^{(\pm)} (x) \, ] = 0$, Schr\"odinger: $i \, \partial_t \, [ \, \tilde{\psi} (x) \, \psi (x) \, ] + \frac{1}{2\, M} \!\stackrel{\rightarrow}{\nabla} \!\cdot \,[ \, \tilde{\psi} (x) \stackrel{\rightarrow}{\nabla} \psi (x) - (\stackrel{\rightarrow}{\nabla} \tilde{\psi} (x)) \,\psi (x) ] = 0$.
Note that all currents are conserved and in general non-zero, even for the neutral KG field! The observed norm conservation for (anti)causal KG \& Dirac fields and Schr\"odinger wavefunctions is related to the (complex) energy conservation and hence related to the probability conservation! We conclude that the Schr\"odinger norm is $\int d^3x \, \tilde{\psi} (x) \, \psi (x)$, and {\em not} $\int d^3x \, |\psi (x)|^2\;$!
\section{Charge conservation in QM, Klein-Gordon- and Dirac-theory}
We will introduce simply charged (anti)causal systems according to the isospin concept. For the KG theory we define $\phi_\pm (x) := \big( \phi_1(x) \pm i\, \phi_2(x) \big)/\sqrt{2}$, for the Dirac theory we define $\psi_\pm (x) \; := \; \big(  \psi_1(x) \pm i\, \psi_2(x)\big)/\sqrt{2}$, while for the Schr\"odinger theory we define $\psi_\pm (x) := \big(  \psi_1(x) \pm i\, \psi_2(x)\big)/\sqrt{2}$ and $\tilde{\psi}_\pm (x) := \big(  \tilde{\psi}_1(x) \pm i\, \tilde{\psi}_2(x)\big)/\sqrt{2}$.\footnote{The neutral theory would follow by setting either $\phi_1(x)$ ($\psi_1(x)$, $\tilde{\psi}_1(x)$) or $\phi_2(x)$ ($\psi_2(x)$, $\tilde{\psi}_2(x)$) equal to zero.} Subtraction of causal equations of motion and the respective adjoints\footnote{KG: $(\partial^2 +M^2) \, \phi_{\pm} (x) = 0$, $\phi_{\pm} (x) \,(\,\stackrel{\leftarrow}{\partial}{}^2 +M^2 ) = 0$; Dirac: $(i\!\not\!\partial - M ) \, \psi_{\pm} (x) = 0$, $\overline{\psi_{\pm}^{\,c}} (x) \, (-\,i\!\stackrel{\leftarrow}{\not\!\partial} - \,M ) = 0$, Schr\"odinger: $+ \, i \,\partial_t \, \psi_{\pm} (x) = (- \,\frac{1}{2\,M} \, \stackrel{\rightarrow}{\nabla} {}^2 + V(x) ) \, \psi_{\pm} (x)$, $- \, i \, \tilde{\psi} (x) \stackrel{\leftarrow}{\partial}_t \, = \, \tilde{\psi} (x) \, (- \, \frac{1}{2\, M} \, \stackrel{\leftarrow}{\nabla} {}^2 + V(x))$.}
 leads to the following continuity equations reflecting charge conservation: \\
\noindent KG: $\partial_\mu [ \, \phi_{\mp} (x) \; \partial^\mu \phi_{\pm} (x)  - (\partial^\mu \phi_{\mp} (x))\,\phi_{\pm} (x) \, ]  = 0$, Dirac: $\partial_\mu [ \, i\; \overline{\psi^{\,c}_{\mp}} (x) \; \gamma^\mu\;  \psi_{\pm} (x) \, ] = 0$, Schr\"odinger: $i \partial_t \, [ \, \tilde{\psi}_\mp (x) \, \psi_\pm (x) \, ] + \frac{1}{2\, M} \stackrel{\rightarrow}{\nabla} \!\cdot \,[ \, \tilde{\psi}_\mp (x) \stackrel{\rightarrow}{\nabla} \psi_\pm (x) - (\stackrel{\rightarrow}{\nabla} \tilde{\psi}_\mp (x)) \,\psi_\pm (x) \, ] = 0$.
Note that currents and charges vanish for neutral fields (as they should)! The reason is a cancellation of underlying norm currents! The neutral Dirac field does {\em not} admit any {\em Abelian} gauge couplings due to $[\,\overline{\psi^{\,c}} (x) \not\!\!A(x) \, \psi (x)\,]^T = \! - \,\overline{\psi^{\,c}} (x) \not\!\!A(x) \, \psi (x)$, $[\,\overline{\psi^{\,c}} (x) \,\sigma^{\mu\nu} \, F_{\mu\nu}(x) \, \psi (x)\,]^T = - \,\overline{\psi^{\,c}} (x) \,\sigma^{\mu\nu} \, F_{\mu\nu}(x) \, \psi (x) $.
The concept of local (non)Abelian gauge invariance in (A)CQT is sketched in the footnote.\footnote{The Dirac Lagrangean with minimally coupled (non)Abelian gauge fields is given by the expression ${\cal L} (x) = \overline{\psi_+^c} (x) \, ( \frac{1}{2} \,\, i \!\! \!\stackrel{\;\,\leftrightarrow}{\not\!\partial} + g \not\!\!A (x) -  M ) \, \psi_- (x) + \overline{\psi}_- (x) \, ( \frac{1}{2} \,\, i \! \!\!\stackrel{\;\,\leftrightarrow}{\not\!\partial} \!\! + g^\ast \gamma^{\,\mu} A^+_{\mu} (x) -  \bar{M} ) \, \psi_+^c (x)$ (with $\psi_\pm (x) := (\psi_1(x)\pm \, i\psi_2(x))/\sqrt{2}$). 
It is invariant under the local gauge transformations $g \not\!\!\!A^{\,\prime} = g \not\!\!\!A + [\not\!\!\partial , \Lambda (x)]$, $\psi^{\,\prime}_- (x) = \exp(+i \, \Lambda (x)) \, \psi_- (x)$, $\psi^{\,\prime}_+ (x) = \exp(- i \, (\Lambda (x))^T) \, \psi_+ (x)$. Non-Abelian case: $A_\mu (x) =  A_\mu^{a} (x) \lambda^a/2$ and $\Lambda (x) =  \Lambda^{a} (x) \lambda^a/2$.
{\em Non-Abelian} gauge fields admit minimal coupling even to neutral Fermions, {\em if} $[A^{\mu}(x)]^T=-A^{\mu}(x)$, as $[\,\overline{\psi^{\,c}} (x) \not\!\! A(x) \, \psi (x)\,]^T = + \,\overline{\psi^{\,c}} (x) \not\!\! A(x) \, \psi (x)$ and $[\,\overline{\psi^{\,c}} (x) \,\sigma^{\mu\nu} \, F_{\mu\nu}(x) \, \psi (x)\,]^T = + \,\overline{\psi^{\,c}} (x) \,\sigma^{\mu\nu} \, F_{\mu\nu}(x) \, \psi (x) $.}
\section{Shadow fields --- the Hermiticity content of the (anti)causal Klein-Gordon- and Dirac-theory}
It is instructive to decompose an (A)CQT into its Hermitian components. Hermitian fields underlying non-Hermitian (anti)causal fields are here called ``shadow fields'' \cite{Kleefeld:Nakanishi:wx,Kleefeld:Stapp:1973aa}\footnote{Note that E.C.G.\ Sudarshan (1972) used the term ``shadow state'' with a different meaning!}. Consider e.g.\ (anti)causal Lagrangeans of neutral (anti)causal spin 0 Bosons and and spin 1/2 Fermions:
\begin{eqnarray} {\cal L}^{\,0}_{\,\phi} (x) & = &  
\frac{1}{2}\, \Big( (\partial \,\phi (x) )^2  - \, M^2 \, (\phi (x))^2 \, \Big)\; + \;
\frac{1}{2}\, \Big( (\partial \,\phi^+ (x) )^2  - \, M^{\ast \, 2} \,
(\phi^+ (x))^2 \Big) \; , \nonumber \\
{\cal L}^{\,0}_{\,\psi} (x) & = & \frac{1}{2} \,\Big( \; \overline{\psi^c} (x) \, ( \frac{1}{2} \, i \! \stackrel{\;\,\leftrightarrow}{\partial\!\!\!/} \! -  M ) \, \psi (x) \; + \; \bar{\psi} (x) \, ( \frac{1}{2} \, i \! \stackrel{\;\,\leftrightarrow}{\partial\!\!\!/} \! -  \bar{M} ) \, \psi^c (x) \; \Big) \; . \label{Kleefeld:EquationLabel29}
\end{eqnarray}
$\phi(x)$, $\phi^+(x)$, $\psi(x)$, $\psi^c(x)$ are decomposed in Hermitian shadow fields $\phi_{(1)}(x)$, $\phi_{(2)}(x)$, $\psi_{(1)}(x)$, $\psi_{(2)}(x)$ by  $\phi(x) =: ( \phi_{(1)}(x) + i \, \phi_{(2)} (x))/\sqrt{2}$, $\phi^+(x) =: ( \phi_{(1)}(x) - i \, \phi_{(2)} (x))/\sqrt{2}$, and $\psi(x) =: ( \psi_{(1)}(x) + i \, \psi_{(2)} (x))/\sqrt{2}$, $\psi^c(x) =: ( \psi_{(1)}(x) - i \, \psi_{(2)} (x))/\sqrt{2}$, yielding the decomposed Lagrangeans
\begin{eqnarray} {\cal L}^{\,0}_{\,\phi} (x) & = & 
\frac{1}{2} \,\Big( (\partial \phi_{(1)} (x) )^2  - \mbox{Re}[M^2]  
(\phi_{(1)} (x))^2 \Big) -  \frac{1}{2} \,\Big( (\partial \phi_{(2)} (x) )^2  -  \mbox{Re}[M^2]  (\phi_{(2)} (x))^2 \Big) \nonumber  \\[1mm] 
 & + & \mbox{Im}[M^2]  \; \phi_{(1)} (x) \,\phi_{(2)} (x) \; , \nonumber \\[2mm] 
{\cal L}^{\,0}_{\,\psi} (x) & = & \frac{1}{2} \, \overline{\psi}_{(1)} (x) \, \Big( \frac{1}{2} \; i \! \stackrel{\;\,\leftrightarrow}{\not\!\partial} \! -  \,\mbox{Re}[M] \Big) \, \psi_{(1)} (x) -  \frac{1}{2} \; \overline{\psi}_{(2)} (x) \, \Big( \frac{1}{2} \; i \! \stackrel{\;\,\leftrightarrow}{\not\!\partial} \! - \, \mbox{Re}[M] \Big) \, \psi_{(2)} (x) \nonumber  \\
 & + &   \frac{1}{2} \; \mbox{Im}[M] \; \Big( \overline{\psi}_{(2)} (x) \;\psi_{(1)} (x) +  \overline{\psi}_{(1)} (x) \;  \psi_{(2)} (x) \Big) \; . \label{Kleefeld:EquationLabel31} 
\end{eqnarray}
Note that Bosonic \& Fermionic shadow fields are described by {\em principal value propagators} and {\em interact} with each other. One shadow field has {\em positive} norm, one has {\em negative} norm. If one would remove the interaction term, one would introduce interactions between causal and anticausal fields (e.g. $\phi(x) \, \phi^+(x)$) leading to a {\em violation of causality} and the {\em loss of analyticity} in QT.

\section{Chiral symmetries in (anti)causal Dirac theory}

\noindent {\bf Chiral rotations of (anti)causal fields.} Consider the (anti)causal Lagrangean given by equation (\ref{Kleefeld:EquationLabel29}).
Perform the following chiral rotation of (anti)causal fields: $\psi (x) \rightarrow \exp (i\,\gamma_5\, \alpha) \, \psi (x)$, $\psi^c (x) \rightarrow \exp (i\,\gamma_5\, \alpha) \, \psi^c (x)$.
The resulting continuity-like equation is $\partial_\mu [ \, \overline{\psi^c} (x) \; \gamma^{\,\mu} \, \gamma_5  \, \psi (x) \, ] \propto M$.
Of course there exist the respective Hermitian conjugate/transposed continuity-like equations!

\noindent {\bf Chiral rotations of shadow fields.} Consider the (anti)causal Lagrangean equation (\ref{Kleefeld:EquationLabel31}).
Perform the following chiral rotation of the shadow fields: $\psi_{(1)} (x) \rightarrow \exp (i\,\gamma_5\, \alpha) \, \psi_{(1)} (x)$, $\psi_{(2)} (x) \rightarrow \exp (-\,i\,\gamma_5\, \alpha) \, \psi_{(2)} (x)$.
The resulting continuity-like equation is $\partial_\mu [ \, \overline{\psi} (x) \; \gamma^{\,\mu} \, \gamma_5  \, \psi (x) \, ] \propto \mbox{Re}[M]$.
Also here there exist the respective Hermitian conjugate/transposed continuity-like equations!

\noindent {\bf Chiral symmetry for massive Fermions.}
Define $\psi_R(x):=P_R \, \psi(x)$,$\, \psi_L(x):=P_L \, \psi(x)$,$\,P_R:=(1+\gamma_5)/2$,$\,P_L:=(1-\gamma_5)/2$. Define also $\chi_\pm (x):= (\psi_R(x) \pm i \psi_L(x))/\sqrt{2}$. Consider the (anti)causal Lagrangean equation (\ref{Kleefeld:EquationLabel29}) in the new fields, i.e.
\begin{eqnarray} {\cal L}^{\,0}_{\,\psi} (x) & = & \frac{1}{2} \Big\{ 
\overline{\psi_L^c} (x) \frac{i}{2} \!\! \stackrel{\;\,\leftrightarrow}{\not\!\partial} \!\! \psi_R (x) + 
\overline{\psi^c_R} (x)  \frac{i}{2} \!\! \stackrel{\;\,\leftrightarrow}{\not\!\partial} \!\! \psi_L (x) - M \Big(
\overline{\psi_R^c} (x)  \, \psi_R (x) + 
\overline{\psi^c_L} (x)  \, \psi_L (x) \Big)
\Big\} + \mbox{h.c.} \nonumber \\[2mm]
 & = & \frac{1}{2} \Big\{ 
\overline{\chi_+^c} (x) \frac{1}{2} \!\! \stackrel{\;\,\leftrightarrow}{\not\!\partial} \!\! \chi_+ (x) - 
\overline{\chi^c_-} (x)  \frac{1}{2} \!\! \stackrel{\;\,\leftrightarrow}{\not\!\partial} \!\! \chi_- (x)  -  M \Big(
\overline{\chi_+^c} (x) \, \chi_- (x) +  
\overline{\chi^c_-} (x) \, \chi_+ (x) \Big) 
\Big\} + \mbox{h.c.}  \nonumber 
\end{eqnarray}
Note that the Lagrangean is invariant under the chiral rotations $\chi_\pm (x) \rightarrow \exp (\pm \,i\,\gamma_5\, \alpha) \, \chi_\pm (x)$ even for {\em arbitrary complex Fermion mass} $M$!
\section{New antiparticle concept and the intrinsic parities of anti-Bosons and anti-Fermions} \label{Kleefeld:section11}
In (A)CQT antiparticles are isospin partners of particles\footnote{Charged pions ($\pi^+$, $\pi^-$) are e.g.\ represented by the isospin combination $\pi_\pm(x) = (\pi_1(x) \pm i \, \pi_2(x))/\sqrt{2}$, while the positron ($e^+$) and electron ($e^-$) are represented by $e_\pm(x) = (e_1(x) \pm i \, e_2(x))/\sqrt{2}$. $\pi_1(x)$, $\pi_2(x)$ or $e_1(x)$, $e_2(x)$ are non-Hermitian fields describing, respectively, pairs of causal neutral particles with equal complex mass.}. This holds for Bosons and Fermions. It leads --- like in the Bosonic case --- to the fact that the anti-Fermions have the same intrinsic parity as the Fermions\footnote{In spite of this feature (A)CQT reproduces exactly
the high precision results of Quantum Electrodynamics.}. The components of Bosonic or Fermionic field operators characterized by phasefactors with ``negative'' (complex)
frequency, i.e. $\exp(-\,i\,\omega(\vec{k})\,t)$ and $\exp(-\,i\,\omega^\ast(\vec{k})\,t)$, are responsible for the annihilation of (anti)particles and respective (anti)holes, and not for their creation. The traditional identification of annihilation operators of negative frequency states with antiparticles gets lost for fields described by a complex mass with a finite imaginary part. In this situation traditional HQT --- in contrast to (A)CQT --- ceases to be applicable. 
\section{Conjugate $T$-Matrix $\overline{T}_{fi}$, dual vacuum,  transition probabilities and (anti)causal cross sections}
As $|\psi(x)|^2$ is not a probability density in (anti)causal Schr\"odinger theory, $|T_{fi}|^2$ is {\em not} to be interpreted as a {\em transition probability} in (anti)causal scattering theory! In (anti)causal scattering theory we have instead to consider a quantity $\overline{T}_{fi} \,T_{fi}$, where $\overline{T}_{fi}$ ($\not= T^+_{fi}$) is called the {\em conjugate $T$-matrix}. The construction of the explicit analytical expression for the conjutate T-matrix $\overline{T}_{fi}$ showed up to be a non-trivial task. For brevity we want to give here the final result without proof. We assume the causal $T$-matrix $T_{fi}$ to be determined by the standard expression $(2\pi)^4 \; \delta^4 (P_f - P_i)\;\; i \; T_{fi} = \left<\!\left<0\right|\right. {\cal A}(\vec{p}^{\,\,\prime}_{\,N_f}) \ldots {\cal A}(\vec{p}^{\,\,\prime}_{\,1}) \; T [ \, \exp(+\,i \, S_{int}) - 1 \, ]\; ({\cal C}(\vec{p}_{\,1}))^+ \ldots ({\cal C}(\vec{p}_{\,N_i}))^+ \left|0\right>_c$
with ${\cal A}(\vec{p}^{\,\,\prime}_{j}) \in \{ a (\vec{p}^{\,\,\prime}_{j}), b (\vec{p}^{\,\,\prime}_{j})\}$ and  ${\cal C}(\vec{p}_{j})\in \{ c (\vec{p}_{j}), d (\vec{p}_{j})\}$. Call $N_F$ the overall number of Fermionic operators in the intitial and final state. Then the conjugate causal $T$-matrix $\overline{T}_{if}$ is given by:
\begin{eqnarray} \lefteqn{(2\pi)^4 \; \delta^4 (P_f - P_i)\; \; (-i) \;\, \overline{T}_{if} \quad =} \nonumber \\
 & = & \left<\!\left<\bar{0}\right|\right. \Big( {\cal A}(\vec{p}_{\,N_i}) \ldots {\cal A}(\vec{p}_{\,1}) \; T [ \, \exp(-\,i \, S_{int}) - 1 \, ]\; ({\cal C}(\vec{p}^{\,\,\prime}_{\,1}))^+ \ldots ({\cal C}(\vec{p}^{\,\,\prime}_{\,N_f}))^+ \Big)^T \left|\bar{0}\right>_c \nonumber \\
 & \stackrel{!}{=} & (-1)^{N_F (N_F-1)/2} \left<\!\left<\bar{0}\right|\right. ({\cal C}(\vec{p}^{\,\,\prime}_{\,N_f}))^+ \ldots ({\cal C}(\vec{p}^{\,\,\prime}_{\,1}))^+ \; \Big( T \,[ \, \exp(-\,i \, S_{int}) - 1 \, ]\Big)^T {\cal A}(\vec{p}_{\,1}) \ldots {\cal A}(\vec{p}_{\,N_i}) \left|\bar{0}\right>_c \nonumber \\
 & = & (-1)^{N_F (N_F-1)/2} \left<\!\left<\bar{0}\right|\right. ({\cal C}(\vec{p}^{\,\,\prime}_{\,N_f}))^+ \ldots ({\cal C}(\vec{p}^{\,\,\prime}_{\,1}))^+ \; \overline{T} \,[ \, \exp(-\,i \, S^{\,T}_{int}) - 1 \, ]\; {\cal A}(\vec{p}_{\,1}) \ldots {\cal A}(\vec{p}_{\,N_i}) \left|\bar{0}\right>_c \nonumber \\
 & \stackrel{!}{=} & (-1)^{N_F (N_F-1)/2} \left<\!\left<\bar{0}\right|\right. ({\cal C}(\vec{p}^{\,\,\prime}_{\,N_f}))^+ \ldots ({\cal C}(\vec{p}^{\,\,\prime}_{\,1}))^+ \; \overline{T} \,[ \, \exp(-\,i \, S_{int}) - 1 \, ]\; {\cal A}(\vec{p}_{\,1}) \ldots {\cal A}(\vec{p}_{\,N_i}) \left|\bar{0}\right>_c \nonumber
\end{eqnarray}
with ${\cal A}(\vec{p}_{j}) \in \{ a (\vec{p}_{j}), b (\vec{p}_{j})\}$ and  ${\cal C}(\vec{p}^{\,\,\prime}_{j})\in \{ c (\vec{p}^{\,\,\prime}_{j}), d (\vec{p}^{\,\,\prime}_{\,j})\}$ and
$\left|\bar{0}\right>$ (and $\left<\!\left<\bar{0}\right|\right.$) being the {\em dual vacuum} annihilating creation operators and creating annihilation operators.
We used the useful transposition identity  
$\big( T\, \big[ \,{\cal O}(x_1) \ldots  {\cal O}(x_n) \, \big] \big)^T \; = \; \overline{T}\, \big[ \,({\cal O}(x_1))^T \ldots  ({\cal O}(x_n))^T \, \big]$ holding for Bosonic and Fermionic operators.
$\overline{T}_{fi} \,T_{fi}$ for a causal process (to be identified with the transition probability) and therefore
also the respective causal cross section are not necessarily real numbers\footnote{Only if the underlying theory represented by a causal Lagrangean or causal Hamiltonian is probability conserving, i.e.\ non-absorptive, $\overline{T}_{fi} \,T_{fi}$ and therefore also the causal cross section will be quasi-real, i.e.\ infinitesimally close to a real number.
 Hence, for selective (so called ``inelastic'') causal processes in probability non-conserving theories (e.g.\ open quantum systems) the respective causal cross sections will develop a finite imaginary part. Within particle physics this new feature complements in a beautiful manner, what is well understood for a long time in theoretical optics, i.e.\ that the imaginary part of the refractive index of a material is related to its absorption coefficient. Similar arguments hold for anticausal processes and respective anticausal cross sections.}. 

\subsection*{Acknowledgements}
This work has been supported by the
{\em Funda\c{c}\~{a}o para a Ci\^{e}ncia e a Tecnologia} \/(FCT) of the {\em Minist\'{e}rio da Ci\^{e}ncia e da Tecnologia (e do Ensinio Superior)} \/of Portugal, under Grants no.\ PRAXIS
XXI/BPD/20186/99, SFRH/BDP/9480/2002, and POCTI/\-FNU/\-49555/\-2002.

\LastPageEnding

\end{document}